# An assessment of the first "scientific habilitation" for university appointments in Italy[1]

*Giovanni Abramo* (corresponding author)
  Laboratory for Studies of Research and Technology Transfer
  Institute for System Analysis and Computer Science (IASI-CNR)
  National Research Council of Italy
  Viale Manzoni 30, 00185 Rome - ITALY
  giovanni.abramo@uniroma2.it
  Tel/fax +39 06 72597362

*Ciriaco Andrea D'Angelo*
  Department of Engineering and Management
  University of Rome "Tor Vergata"
  Via del Politecnico 1, 00133 Rome - ITALY
  dangelo@dii.uniroma2.it

## Abstract

Nations with non-competitive higher education systems and with high levels of corruption, are more exposed to phenomena of discrimination and favoritism in faculty recruitment. Italy is a case in point, as shown by empirical studies, judicial reports and media attention. Governments have intervened repeatedly to reduce the problem, with scarce success. The 2010 reforms to the university recruitment system provided that access to the ranks of associate and full professor would now be possible only through an initial "scientific habilitation " to be awarded by sectorial committees of national experts. The objective of this work is to analyze the relationship of the recent habilitation procedure outcomes to the actual scientific merit of the various candidates, as well as to other variables that are explicative of possible practices of favoritism and discrimination. The analyses identify the presence of potential cases of discrimination and favoritism.

## Keywords

*Research evaluation; recruitment; bibliometrics; Italy*

## JEL codes

I230; O380

---

# 1. Introduction

In the current knowledge-based economy, the quality of human capital is key in determining the competitive advantage of economic systems. Universities, as the institutions dedicated to the education of human capital, thus assume ever more vital importance for global competitiveness and socio-economic development. The strengthening of the higher education systems has thus become increasingly central in the policy agenda of a growing number of countries. The effectiveness of recruitment and advancement of university professors is crucial to the search for excellence. In competitive higher education systems (such as in the U.S. and the U.K.) the pursuit of competitive advantage leads to the development of world-class universities, whose main distinctive competence is their ability to attract, develop and retain highly-talented national and foreign faculty. In those systems, appointments to academic positions are handled through local *ad hoc* search committees or advertisements in international social network and scientific journals. Differently, in several European countries, including Germany, Sweden, Denmark and Italy, where the state has a significant presence in the regulation and direct financing of universities, competitive mechanisms are often weak, and faculty recruitment and advancement take place by rigid procedures, frequently regulated by a central bureaucracy (Auranen and Nieminen, 2010). Moreover, nations where non-competitive higher education systems are further associated with high levels of corruption, are more exposed to phenomena of favoritism and discrimination in faculty recruitment and career advancement. Italy is a case in point. According to The Global Competitiveness Report 2014-2015 (Schwab, 2014), Italy ranks $106^{th}$ out of 144 countries in deterioration in the functioning of its institutions, and $134^{th}$ in favoritism in decisions of government officials; while placing $54^{th}$ out of 150 countries in the 2014 World Democracy Audit for corruption.[2] It is no surprise then that the Italian *concorso,* nationally regulated competition for academic recruitment, is generally perceived as an unfair selection practice, as shown by empirical studies, judicial reports and media attention (Zagaria, 2007; Perotti, 2008). Various Italian governments have made countless attempts to overcome the chronic social disease of favoritism, by changing the rules and procedures of the academic appointment system, however with scarce success. Most recently, hopes were raised in many parts of the academic community when an announcement emerged that career-path competitions would now be based on quantitative indicators. In fact new legislation, introduced in 2010, provided that the access to faculty positions would take place through competitions announced by each individual university, and open exclusively to the participation of candidates holding a so-called national "scientific habilitation". The habilitation process was to involve evaluations of the candidates by committees of experts, relevant to the many fields of the applicants' scientific activity.

    Habilitation is not a new mechanism in the panorama of European university recruitment processes. For a number of years, it has also been applied in other forms in Germany (Enders, 2001), and more recently in France (Musselin, 2004) and Spain (MEC, 2007). A new and unique feature of the Italian habilitation is that it involves the use of bibliometric indicators to support the committee evaluations, in this case with the data on the indicators provided by the Italian Ministry for Education, University and Research (MIUR). In fact in the ministry's initial scheme, any award of habilitation was

---

[2] http://www.worldaudit.org/corruption.htm, last accessed on 01/07/2015.

to be subsequent to passing the threshold values in all the three bibliometric indicators of research performance, thus creating an outright block against mediocre candidates arriving in the rank of professor.

However the publication of the decree spelling out the evaluation criteria and parameters, followed by the release of the threshold values, unleashed an increasingly heated debate. Those who were favorable recognized the measures as new guarantees of greater objectivity and transparency, imposed over the familiar situation. Those who were against, stigmatized the use of bibliometric criteria for the evaluation of scientific activity, which requires evaluation by peer review. In the end the "opposers" won, in spite of the fact that many studies had already demonstrated a high correlation between the value of scientific production assessed by bibliometric indicators, and that assessed through peer review processes (Wainer and Vieira, 2013; Li et al., 2010; Norris and Oppenheim, 2010; Reinhart, 2009; Meho and Sonnenwald, 2000). The Minister himself (who had by that time replaced the original sponsor of the changes) had to intervene to clarify that the expert committees would be paramount in defining the parameters and criteria for evaluating the candidates in their sectors, and that the ministerial decree simply set "guidelines": the candidates' standing in respect to the median of the ministerial indicators was neither a condition (nor any guarantee) of obtaining habilitation. This sequence of decisions opened notable breaches in the dyke of quantitative criteria that had been introduced to hold back the generalized practices of favoritism. As was to be expected, the publication in 2013 of the results of the first habilitation again unleased a storm of polemics, supported by troubling analyses, case stories and anecdotes. The most startling paradox that emerges from our investigation is that of three candidates habilitated by the same committee to full professors, but not to associate professors. In his 2015 annual address, the president of the Region of Lazio Administrative Court, reported that 1240 proceedings had been launched by the candidates of habilitation competitions[3]. After a second habilitation round, all this led the government to interrupt the procedure for the subsequent habilitation competitions, to make important corrections. It also led the authors to undertake the study presented in this manuscript.

The first Italian habilitation competition has been so far the object of two other studies. Marzolla (2015) computed descriptive statistics providing a picture of the outcome of the qualification procedure. He also analysed the correlation between outcome and the values of bibliometric indicators. Pautasso (2015) instead explored gender issues in both participation and outcomes. The object of the present work is the analysis of the relation between the outcomes of the national habilitation, the scientific merit of the candidates (measured by the values of the bibliometric indicators), and other explicative variables of possible practices of favoritism and discrimination. The diagnoses that emerge from these analyses could be of further support to the government as it readies its corrective measures for the next habilitation competitions. In a longer-term and international view, the aim of the investigation is also to examine the potential of bibliometrics not only as a support for the comparative evaluation of individual candidates, but also as an instrument for the ex-post assessment of the work by evaluation committees. In this sense, bibliometrics could serve not only towards the efficient selection of faculty, but also as a deterrent to practices of discrimination and favoritism in selection processes.

---

[3] http://roma.repubblica.it/cronaca/2015/02/26/news/tar_lazio_presidente_2014_aumento_consistente_ controversie-108226685/ last accessed on 01/07/2015.

Indeed, problems of fairness in academic appointments are not limited to Italy. The international literature has dedicated considerable attention to the study of academic recruitment and promotion, largely regarding questions of gender and minority discrimination (Zinovyeva and Bagues, 2015; van den Brink et al., 2010; Cora-Bramble, 2006; Price et al., 2005; Trotman et al., 2002; Stanley et al., 2007). One of the conclusions is that discriminatory phenomena seem to develop above all when evaluations are based on non-transparent criteria (Rees, 2004; Ziegler, 2001; Husu, 2000; Ledwith and Manfredi, 2000; Allen, 1988). In effect, academic recruitment is often reported as an informal process in which a few powerful professors select new ones through cooptation mechanisms (van den Brink et al., 2010; Husu, 2000; Fogelberg et al., 1999; Evans, 1995). Such mechanisms often conceal the phenomenon of favoritism, which has been intensively examined in only a few nations, such as Turkey (Aydogan, 2012), Australia (Martin, 2009), Spain (Zinovyeva and Bagues, 2015) and Italy (Abramo et al., 2014a; 2014b; Perotti, 2008; Zagaria, 2007). In Zinovyeva and Bagues' (2015) examination of the Spanish university system, the investigators concentrated on the role of connections between the candidates and the evaluators composing the examining boards deciding on academic promotions. They show that the future performance of candidates who were promoted and had a weak connection with the evaluators was better than that of their colleagues. Conversely, successful candidates with a strong link to the evaluators register worse performance both before and after their promotion.

For Italy, Abramo et al. (2014a) investigated the effectiveness of the selection process for academic advancement, referring to the case of 287 associate professor competitions launched in 2008. The analysis showed that the new associate professors were on average more scientifically productive than their incumbent colleagues. However several critical issues appeared, particularly concerning unsuccessful candidates who outperformed the competition winners in terms of research productivity, as well as a number of competition winners who resulted as totally unproductive. Almost half of individual competitions selected candidates who would go on to achieve below-median productivity in their field of research over the subsequent triennium.

In a further work, the same authors attempted to discover the potential factors that could have contributed to the outcomes of the 2008 round of Italian competitions (Abramo et al., 2014b). They identified that the main determinant of a candidate's success was not their scientific merit, but rather the number of their years of service in the same university as the committee president. Where the candidate had actually cooperated in joint research work with the president, the probability of success again increased significantly. These results confirm what Zinovyeva and Bagues (2015) have demonstrated for the Spanish case, i.e. the existence of mechanisms of selection that are completely contrary to the intentions of the regulatory framework for the competitions.

Before presenting our work, it must be acknowledged from the very beginning that research performance is not the only dimension of quality of a candidate to academic positions. In addition, the bibliometric indicators introduced by the legislator in any case represent a proxy of the real value of a researcher's scientific activity. The bibliometric assessment of research performance by quantity and quality of output neglects other attributes of the scientists' activities, for example the ability to manage research teams, to attract funds, their activities in consulting, teaching, editorial work, technology transfer, and so on. For this, other than the bibliometric criteria, the normative framework for the habilitations indicated a list of further criteria that the committees could consider, and left it to the committees themselves to explicitly define the evaluation model that they would apply. Still, common sense would lead us to believe that there is a strong correlation between research productivity and all other dimensions of scientific merit.

The next section of the paper summarizes the overall characteristics of the Italian academic system, while Section 3 provides greater detail on the "scientific habilitation" process. Section 4 describes the results of the habilitation competitions by bibliometric sector and provides a regression analysis to determine the impact of the possible determinants. Section 5 attempts to quantify the cases of apparent discrimination and favoritism by sector of competition. Section 6 concludes the work with the authors' discussion and recommendations.

## 2. The Italian higher education system and recruitment of university professors

The MIUR recognizes a total of 96 universities as having the authority to issue legally-recognized degrees. Sixty-seven (70%) are public, but employ 95% of the overall faculty. Universities are largely financed by government through non-competitive allocations. It was only following the first national research evaluation exercise (VTR) conducted between 2004 and 2006, that a minimal share, less than 4% of total income, was assigned by the MIUR on the basis of the assessment of research. This share has currently been raised to 18% of total allocations.

In keeping with the Humboldtian model, there are no "teaching-only" universities in Italy, and all professors are required to carry out both research and teaching. National legislation includes a provision that each faculty member must provide a minimum of 350 hours per year of teaching. At the close of 2014, there were 55,672 faculty members and a roughly equal number of technical-administrative staff. All new personnel enter the university system through public competitions, and career advancement can only proceed by further public competitions. Salaries are regulated at the centralized level and are calculated according to role (administrative, technical, or professorial), rank within role (for example assistant, associate or full professor) and seniority. None of a professor's salary depends on merit. Moreover, as in all Italian public administration, dismissal of unproductive employees is unheard of.

The entire legislative-administrative context creates an environment and culture that are barely competitive, yet flourishing with favoritism and other opportunistic behaviors that are dysfunctional to the social and economic roles of the higher education system. The overall result is a system of universities that are almost completely undifferentiated for quality and prestige, with the exception of the three tiny Schools for Advanced Studies and a very small number of private, special-focus universities (Abramo et al., 2012). The system is thus unable to attract significant talented foreign faculty, or even students. The numbers are negligible: only 1.8% of research staff are foreign nationals. Over the 2004-2008 period, 6,640 (16.8%) of the 39,512 hard-sciences professors did not publish any scientific articles in the journals indexed by the Web of Science (WoS). Another 3,070 professors (7.8%) did achieve publication, but their work was never cited (Abramo et al., 2011). This means that 9,710 individuals (24.6%) had no impact on scientific progress measurable by bibliometric databases.[4] An almost equal 23% of professors alone produced 77% of the overall Italian scientific advancement. Differently from competitive higher education systems, this 23% of "top" faculty is not concentrated in a limited number of universities, but is instead dispersed more or less uniformly among all Italian universities, along with the unproductive academics, so that no single institution reaches the critical mass of excellence necessary to develop as an elite university and compete at the international level (Abramo and D'Angelo, 2014).

The recruitment and career advancement of professors are regulated by specific law, overseen by the MIUR. In the attempt to reduce favoritism in recruitment, there have been major reforms of the norms over recent years, with the last one being Law 240 of 2010, which introduced a double level of evaluation for selection of associate and full professors. The first level is national, managed directly by the MIUR, and is intended to habilitate all candidates who have sufficient scientific qualifications. The second is managed by the individual universities, to select the candidates best suited to the specific needs of the university from among those first habilitated at the national level.

Prior to Law 240, the processes of recruitment and career advancement were in the hands of the individual universities, following procedures dictated at the central level. The transparency provisions, the nomination of a national committee of experts in the field, and the timely issue of regulations for the evaluation procedures were all intended to ensure effectiveness in the selection process. In reality, the characteristics of Italian system - such as the generally strong inclination to favoritism, the structured absence of responsibility for poor performance by research units, and the lack of incentive schemes for merit - undermined the credibility of selection procedures for hiring and advancement of university personnel, just as happens for the Italian public administration in general. The intention of the policy makers in introducing the habilitation competition prior to the actual recruitment was to reverse past trends and to limit or even eliminate the inherent inefficiencies and favoritism. The "scientific habilitation", through the objective evaluation of the candidates' scientific merit, was to guarantee the appropriate pre-selection of those then eligible to apply for competitions launched by the individual universities.

## 3. The scientific habilitation process

The regulations governing the national scientific habilitation descend from Law 240 of 2010[5]. and are detailed by: "Regulation concerning the awarding of national scientific habilitation", contained in Presidential decree 222 of 14/09/2011 and "Regulation specifying criteria and parameters for the evaluation of the candidates and the assurance of Committee-members qualifications", in MIUR Ministerial decree 76 of 07/06/2012. The MIUR also created a new Agency for the Evaluation of University and Research Systems (ANVUR), which was assigned responsibility for certain aspects of implementing the habilitation.

According to the regulatory framework, every two years the MIUR must issue a call for appointments to 184 "judging committees", with one committee for each competition sector (CS). The CSs derive from the aggregation of the so-called Scientific Disciplinary Sectors (SDSs, 370 in all), which serve in general for the classification and administration of all Italian university faculty: under this system, each professor is unequivocally classified as practicing in a specific SDS. The SDSs, and so also their aggregations in CSs, are grouped in 14 University Disciplinary Areas (UDAs). Of the 184 CSs, ANVUR classifies 109 as "bibliometric", and the remaining 75 as "non-bibliometric". The call for appointments to the judging committees is open to all full professors categorized in the particular CS. In the most recent call, professors seeking appointment were to present an application including indications of their scientific production: from this, ANVUR calculated three specific indicators (which were different for the bibliometric and non-bibliometric CSs). The request for appointment was eligible if all the indicators showed values above the median for the distribution of all full professors in the CS,[6] otherwise it was immediately rejected. The committee of four Italian academics was then formed by random draw from all the candidates that passed the bibliometric test. The Italian committee members were also joined by a fifth member, selected at random from a list of academics working in OECD nations, whose candidatures had again been screened by ANVUR.

---

[4] Such "unproductive" researchers could publish in journals not indexed by WoS, or codify new knowledge in other forms, such as books, patents, etc.
[5] Law 240 of 2010 was in reality a broad reform of the university system, covering academic recruitment as one part.

The call for candidates to the habilitation is instead issued on an annual basis. Applicants must apply for specific CSs and ranks, however there is no limit on the number of CSs requested, and applicants can also request habilitation to both associate and full professor at the same time. For the 2012 call, there were also no minimum requirements for admissibility to the competitions. The candidatures were to be accompanied by a curriculum vitae and a list of the individual's degrees and scientific publications. The latter served for the calculation of three indicators. For the "non-bibliometric" CSs, the same indicators were applied for the evaluation of the candidates to habilitation and for the applicants to committee positions: i) the number of monographs (only those with ISBN); ii) the number of articles in a set of journals indicated by ANVUR, or as part of monographs with ISBN; iii) the number of articles in top-ranked journals, as selected by ANVUR.

For the "bibliometric" CSs the indicators were not identical, but still similar to those used for the selection of the committee members. For the candidates to habilitation, they were:[7]

- number of articles in journals over the period 2002-2012, with normalization in the case of academic "age" less than 10 years (calculated beginning from the first year of publication);
- number of citations received for the overall scientific production, normalized for the academic age of the candidate;
- contemporary h-index (Sidiropoulos et al., 2007) of the overall scientific production.

The above indicators were calculated on the basis of the publications indexed in two major commercial bibliometric sources: Scopus and Web of Science (WoS). In the case of publications indexed in both sources, the value of citations assigned was the higher of the two.

---

[6] The database of the publications used to calculate the median for Italian professors is not made available to the public, but contains only publications voluntarily inserted by any and all individual professors of Italian universities.

[7] For the applicants to the committee-member positions, judging was on the quantity and impact of production over their entire careers, without any normalization for years of academic service.

Beginning from a database of all the publications entered by Italian professors over the course of 2012, ANVUR proceeded to calculate the different indicators for full and associate professors and then published the median values, to serve in the selection of both the committee-members and their subsequent evaluations of the candidates.

Table 1 provides the descriptive statistics for the 2012 call for habilitations. The response was over 59,148 applications, more than one sixth of which were for Medicine (9,987), followed by Ancient history, philology, literature, art (6,324 applications, 10.7% of total) and Biology (6,244 applications, 10.6% of total). At the bottom of the list was Earth sciences, with only 1,231, for 2.1% of the total. The requests for full professor habilitation were 30.5% of the total: the detail per UDA shows the maximum incidence was in Mathematics and computer science (36.6%), followed by Economics and statistics (36.2%). The lowest incidence of requests for full professor habilitation was in Political and social sciences (24.2%).

*Table 1: Statistics for the 2012 Italian scientific habilitation (elaborated by the authors from data published on the site for the habilitation - http://abilitazione.miur.it/public/candidati.php?sersel=50&)**

| UDA | CS | Of which "bibliometric" | Applications | Of which for full professor |
|---|---|---|---|---|
| 01 - Mathematics and computer science (MAT) | 7 | 7 | 2,492 | 911 (36.6%) |
| 02 - Physics (PHY) | 6 | 6 | 4,372 | 1,451 (33.2%) |
| 03 - Chemistry (CHE) | 8 | 8 | 2,344 | 695 (29.7%) |
| 04 - Earth sciences (EAR) | 4 | 4 | 1,231 | 400 (32.5%) |
| 05 - Biology (BIO) | 13 | 13 | 6,244 | 1,690 (27.1%) |
| 06 - Medicine (MED) | 26 | 26 | 9,987 | 3,298 (33.0%) |
| 07 - Agricultural and veterinary sciences (AVS) | 14 | 14 | 2,093 | 650 (31.1%) |
| 08 - Civil engineering and architecture (CE) | 12 | 7 | 3,599 | 1,027 (28.5%) |
| 09 - Industrial and information engineering (IIE) | 20 | 20 | 4,535 | 1,573 (34.7%) |
| 10 - Ancient history, philology, literature, art (LIT) | 19 | 0 | 6,324 | 1,718 (27.2%) |
| 11 - History, philosophy, pedagogy, psychology (PP) | 17 | 4 | 5,909 | 1,491 (25.2%) |
| 12 - Law (LAW) | 16 | 0 | 3,037 | 887 (29.2%) |
| 13 - Economics and statistics (ECS) | 15 | 0 | 4,853 | 1,755 (36.2%) |
| 14 - Political and social sciences (POL) | 7 | 0 | 2,128 | 515 (24.2%) |
| Total | 184 | 109 | 59,148 | 18,061 (30.5%) |

*The regulations dictated that after 60 days from the publication of the committee decisions, the data would be obscured. The site now lists only the successfully habilitated candidates.*

To understand the scope of this evaluation procedure, Table 2 indicates the numerosity of the applicants per UDA in relation to the national academic staff in the same UDA. The 59,148 applications were presented by 40,228 candidates. There were an average of 1.47 applications per candidate: 72% presented only one application; 18% presented two; 8% presented between 3 and 5 applications, and 1% presented more than 5. Each candidate on average applied for 1.34 CSs (3,714 candidates applied to both associate and full professor positions in the same CS; 818 of them in more than one CS).

Over half the candidates (55.7%) were already on faculty staff at an Italian university[8]. Physics was the UDA "most open" to candidatures from outside the

---

[8] To isolate all those candidates who were already on staff at national universities, the data downloaded from the habilitation website were crossed with those from the MIUR database on teaching staff (cercauniversita.cineca.it/php5/docenti/cerca.php, last accessed on 01/07/2015). An algorithm for

university sphere: only a third of the applications came from candidates already on faculty. This is traditionally a strong area of the national scientific system, as also witnessed by the significant numbers and dimensions of physics research institutes outside the university system. Many scientists in such research institutes saw the "habilitation" as an opportunity to acquire a form of additional qualification for their internal career progression, or for the option of a potential shift to the academic sphere. On the opposite front, in Economics and statistics and Mathematics and computer science, over two thirds of the total applicants were already on faculty in Italian universities.

Applicants outnumber full and associate professors already on staff in the total Italian faculty. The applications exceed the numbers of the existing staff in all UDAS except in Law: in Physics the 3,122 applicants are over twice the full and associate professors already on staff. Of the existing assistant and associate professors, 52.3% applied for habilitation, with a peak in Political and social sciences (66.1%), followed by Ancient history, philology, literature, art (65.9%) and by Physics (64.9%). The lowest percentage was registered in Chemistry (47.1%).

*Table 2: Incidence of applicants in total national research staff*

| UDA | Total applicants (a) | Of which on faculty (b) | Total Associate + Full professors (c) | Total Assistant + Associate professors (d) | a/c (%) | b/d (%) |
|---|---|---|---|---|---|---|
| 01-MAT | 2,072 | 1,386 (66.9%) | 1,827 | 2,293 | 113.4 | 60.4 |
| 02-PHY | 3,122 | 1,110 (35.6%) | 1,264 | 1,711 | 247.0 | 64.9 |
| 03-CHE | 1,695 | 1,082 (63.8%) | 1,513 | 2,295 | 112.0 | 47.1 |
| 04-EAR | 973 | 404 (41.5%) | 565 | 828 | 172.2 | 48.8 |
| 05-BIO | 3,961 | 1,905 (48.1%) | 2,385 | 3,769 | 166.1 | 50.5 |
| 06-MED | 7,354 | 3,798 (51.6%) | 4,799 | 7,845 | 153.2 | 48.4 |
| 07-AVS | 1,811 | 1,100 (60.7%) | 1,613 | 2,295 | 112.3 | 47.9 |
| 08-CE | 3,075 | 1,690 (55.0%) | 1,938 | 2,686 | 158.7 | 62.9 |
| 09-IIE | 3,561 | 2,171 (61.0%) | 3,015 | 3,800 | 118.1 | 57.1 |
| 10-LIT | 5,126 | 2,565 (50.0%) | 2,818 | 3,891 | 181.9 | 65.9 |
| 11-PP | 4,442 | 2,157 (48.6%) | 2,640 | 3,467 | 168.3 | 62.2 |
| 12-LAW | 2,428 | 1,580 (65.1%) | 2,667 | 3,285 | 91.0 | 48.1 |
| 13-ECS | 2,922 | 1,966 (67.3%) | 2,793 | 3,317 | 104.6 | 59.3 |
| 14-POL | 1,665 | 8,76 (52.6%) | 877 | 1,326 | 189.9 | 66.1 |
| Total | 40,228 | 22,404 (55.7%) | 30,714 | 42,808 | 131.0 | 52.3 |

## 4. Analyses of the 2012 habilitation results

Our analysis is limited to the "bibliometric" CSs alone, because these are the ones where the use of bibliometric parameters and indicators, such as those applied in the habilitation, permits reliability and robustness in the evaluation of the candidates. There are two UDAs that include both bibliometric and non-bibliometric CSs: Civil engineering and architecture, and History, philosophy, pedagogy, psychology. For these UDAs our analysis refers only to the bibliometric CSs alone. For convenience we refer to these UDAs by the acronyms CE (Civil engineering) and PP (Pedagogy and psychology), after the names of the bibliometric CSs contained.

---

disambiguation was applied to resolve cases of homonymy. In several hundred cases the disambiguation was verified manually.

From the open-access habilitation website[9] we have downloaded the data for each candidate (family and first names; sector(s) of competition; full and/or associate professor habilitation sought). For each individual we have also downloaded the values of the bibliometric indicators measured by MIUR from the scientific production declared, as well as the final result of the evaluation (habilitated, not habilitated). From the website we have also taken the national medians for the indicators used, and the composition of the examining committees.

Roughly 36,000 applications for habilitation arrived, referring to 109 bibliometric sectors, and 60% of total applications. Less than a third of the applications are for habilitation to full professor, with the maximum value (36.6%) in Mathematics and the minimum (26.0%) in Pedagogy and psychology.

Of the total applications, 51.4% were presented by assistant and associate professors already on staff at Italian universities, with a maximum in Mathematics (67.2%) and a minimum in Physics (37%). The applications by incumbents are particularly concentrated on habilitation to full professor (67.4% in total) with a peak in Civil engineering, at 84.4%. The applications for habilitation to associate professor are 43.9% of total, with a maximum in Mathematics (60%) and a minimum in Physics (31.3%).

Habilitation was awarded to 44.6% of the candidates, with identical incidence for both academic ranks considered (Table 3). The detail by UDA reveals that in reality the percentage habilitated varies greatly between the disciplines: the minimum value occurs in Pedagogy and psychology (27.5% for full professors and 35.4% for associates); this is also the UDA where the difference between the two percentages is greatest. In Chemistry, the percentage of habilitated candidates is particularly high, at 55.7% for the lower rank and 56.6% for the higher one. Agricultural and veterinary sciences features the highest percentage of candidates habilitated to full professor (57.4%), compared to 51.8% of candidates being habilitated to associate. Exactly the opposite balance occurs in Earth sciences: 37% of the candidates for full professor are habilitated, against 44.2% for the candidates to associate. Isolating those applications for habilitation presented by candidates already in the national faculty, we note that in general there is a higher percentage of habilitations awarded (49.7% for candidates to full professor, 57.0% for those to associate) than we see among the applicants that are not on faculty (34.1% for full professor and 34.9% for associate). Among the "incumbent" candidates the peak of incidence for habilitation to full professor occurs in Agricultural and veterinary sciences (61.7%); for habilitation to associate professor the peak is in Physics (70.8%). The minimum is in Pedagogy and psychology, both for the full professor (36.9%) and associate professor (49.9%) habilitation. The success rate for non-faculty applicants shows a minimum for candidates to full professor in Pedagogy and psychology (5.4%), and in Civil engineering (19.3%) for candidates to associate professor.

---

[9] http://abilitazione.miur.it/public/pubblicarisultati.php, last accessed 01/07/2015. Unfortunately the regulations dictated that after 60 days from the publication of the committee decisions, the data would be obscured. The site now lists only the successfully accredited candidates.

*Table 3: Rates of success per UDA, academic rank and type of applicant*

|  | Habilitated total (%) | | Habilitated among faculty (%) | | Habilitated non-faculty (%) | |
|---|---|---|---|---|---|---|
| UDA | Full | Associate | Full | Associate | Full | Associate |
| 01-MAT | 39.1 | 45.2 | 42.8 | 55.4 | 24.5 | 29.9 |
| 02-PHY | 52.4 | 57.4 | 58.9 | 70.8 | 46.2 | 51.3 |
| 03-CHE | 55.7 | 56.6 | 58.7 | 69.4 | 45.6 | 40.8 |
| 04-EAR | 37.0 | 44.2 | 47.4 | 65.4 | 24.9 | 32.7 |
| 05-BIO | 45.1 | 41.2 | 48.5 | 53.1 | 39.8 | 34.2 |
| 06-MED | 41.8 | 39.9 | 46.1 | 46.3 | 33.1 | 35.1 |
| 07-AVS | 57.4 | 51.8 | 61.7 | 67.3 | 42.4 | 35.0 |
| 08-CE | 36.4 | 39.2 | 40.0 | 55.0 | 16.7 | 19.3 |
| 09-IIE | 44.0 | 42.4 | 52.9 | 63.2 | 15.7 | 21.4 |
| 11-PP | 27.5 | 35.4 | 36.9 | 49.9 | 5.4 | 24.0 |
| Total | 44.6 | 44.6 | 49.7 | 57.0 | 34.1 | 34.9 |

The rate of success for assistant and associate professors already on staff at Italian universities was thus higher than for non-academic candidates. To verify if this corresponded to an equally better scientific profile, we conduct a statistical analysis relating the probability of success of the candidates with the only variables that we can measure for all the candidates: their being already (or not) on faculty and their bibliometric performance. For the statistical model we choose the logistic regression function (rendered linear through the *logit* function), which is particularly suited for modelling dichotomous dependent variables. Formally, the statistical model is described:

$$logit(p) = \beta_0 + \beta_1 X_1 + \beta_2 X_2$$

[1]

Where:

$$logit(p) = log \frac{p(E)}{1-p(E)}$$

$E$ = 1 if the applicant was habilitated; otherwise 0;
$p(E)$ = probability of event E;
$\beta$ = generic regression coefficient;
$X_1$ = applicant's bibliometric performance;
$X_2$ = 1 if the applicant was already on faculty as of 31/12/2012; otherwise 0.

The explicative variable $X_1$ equals the average of the normalized values of each indicator.[10] Values of each indicator are normalized to the median of the distribution of professors of the rank and the CS to which the candidate applied. We give the example calculation of a candidate for habilitation to full professor in CS 09/B3 (Engineering and management). Table 4 illustrates the calculation $X_1$ (last row). In this case the performance $X_1$ of the candidate is 3.56 (average of the ratios of the candidate's values for the three indicators to the relevant medians of the distribution of all full professors in CS 09/B3).

---

[10] Our averaging of the indicators implies the assumption that each had equal importance and impact on the final decision by the committee. However we observe that the Spearman index of correlation between the second and third indicators is 0.94. This could have induced the committee members to give greater relative weight to the first indicator. Still, in the minutes of the meetings where the evaluation criteria were declared, the committees give no indication of any such differentiation, therefore our assumption of equal arithmetic weight appears the correct one.

*Table 4: Example calculation of bibliometric performance based on the three indicators used in the habilitation*

| Index | Value for the candidate | Median for full professors in 09/B3 | Ratio |
|---|---|---|---|
| 1 - Articles in journals over years 2002-2012 | 43 | 7 | 6.14 |
| 2 - Normalized citations | 17.82 | 7 | 2.55 |
| 3 - Contemporary h-index | 10 | 5 | 2.00 |
| | | $X_1 =$ | 3.56 |

Table 5 presents the result of the regressions, aggregating the observations by UDA, for the habilitation to full professor. The p-values for z-test indicate that in all the UDAs, the second variable ($X_2$) is significant. We recall that the value of the Odds Ratio, OR (i.e. $e^\beta$) equals 1 when the associated explanatory variable has no effect on the dependent variable. In fact in all the UDAs, the OR associated with $X_2$ is well over 1, with a peak in Pedagogy and psychology (10.529) and also a very high value in Industrial and information engineering (6.906). However $X_1$ results as non-significant in UDAs 1, 7, 8 and 11. In these UDAs the committees must have adopted additional evaluation criteria whose weight was evidently higher than given bibliometric indicators. The comparison of the standardized coefficients ($\beta_{1Std}$ and $\beta_{2Std}$)[11], shows that there are only three UDAs (Chemistry, Biology and Medicine) where we can affirm with certainty that bibliometric performance has greater influence than being "incumbent" on the success of a candidate. In the other UDAs the scarce explanatory power of the estimates does not allow to reach any conclusions on the determinant of success of a candidate.

*Table 5: Logistic regression results predicting habilitation outcomes for full professor*

| UDA | Obs | Log pseudo likelihood | PseudoR² | $X_1$ (Bibliometric performance) | | | | $X_2$ (Incumbent vs newcomer) | | | |
|---|---|---|---|---|---|---|---|---|---|---|---|
| | | | | $\beta_1$ | p>\|z\| | OR | $\beta_{1Std}$ | $\beta_2$ | p>\|z\| | OR | $\beta_{2Std}$ |
| 01-MAT | 911 | -598.41 | 0.018 | -0.013 | 0.255 | 0.987 | -0.060 | 0.822*** | 0 | 2.274 | 0.330 |
| 02-PHY | 1,451 | -985.18 | 0.019 | 0.145*** | 0.006 | 1.156 | 0.210 | 0.526*** | 0 | 1.692 | 0.263 |
| 03-CHE | 695 | -423.19 | 0.113 | 1.372*** | 0 | 3.942 | 1.325 | 0.645*** | 0.001 | 1.906 | 0.271 |
| 04-EAR | 400 | -239.77 | 0.090 | 0.373*** | 0 | 1.452 | 0.577 | 1.198*** | 0 | 3.315 | 0.598 |
| 05-BIO | 1,690 | -1080.12 | 0.072 | 0.784*** | 0 | 2.191 | 0.746 | 0.380*** | 0 | 1.462 | 0.185 |
| 06-MED | 3,298 | -2069.86 | 0.076 | 0.534*** | 0 | 1.706 | 0.813 | 0.642*** | 0 | 1.901 | 0.303 |
| 07-AVS | 650 | -432.23 | 0.025 | 0.026 | 0.426 | 1.026 | 0.244 | 0.797*** | 0 | 2.218 | 0.331 |
| 08-CE | 385 | -242.70 | 0.038 | 0.088 | 0.196 | 1.092 | 0.301 | 1.358*** | 0.001 | 3.888 | 0.493 |
| 09-IIE | 1,573 | -971.03 | 0.100 | 0.223** | 0.013 | 1.250 | 0.436 | 1.932*** | 0 | 6.906 | 0.825 |
| 11-PP | 374 | -194.92 | 0.115 | 0.076 | 0.136 | 1.079 | 0.213 | 2.354*** | 0 | 10.529 | 1.077 |

*Dependent variable: habilitation outcome; method of estimation: logistic regression QMLE; $\beta$ = raw coefficient; $\beta_{Std}$= X standardized coefficient. OR= Odds Ratio (exp β). z = z-score for test of β=0; p>|z| = p-value for z-test; Statistical significance: \* p-value <0.10, \*\* p-value <0.05, \*\*\* p-value <0.01.*

The results of the analyses concerning the habilitation to associate professors are presented in Table 6. The regressors are significant in all UDAs except in Pedagogy and psychology for $X_1$. For variable $X_2$, the lowest OR is observed in Medicine (1.729). In all the other UDAs, OR is greater than 2, and in three UDAs (Agriculture and veterinary sciences, Civil engineering, Industrial and information engineering) it is even greater than 4. The comparison between $\beta_{1Std}$ and $\beta_{2Std}$ indicates that in Mathematics, Physics,

---

[11] We can compare the effects of these two variables only through the standardized coefficients, since $X_1$ and $X_2$ are measured in different metrics.

Earth sciences, Civil engineering and Industrial engineering, the fact of being incumbent had greater impact than scientific performance in determining the candidate's success.

*Table 6: Logistic regression results predicting habilitation outcomes for associate professor*

| UDA | Obs | Log pseudo likelihood | PseudoR$^2$ | X$_1$ (bibliometric performance) | | | | X$_2$ (incumbent vs new comer) | | | |
|---|---|---|---|---|---|---|---|---|---|---|---|
| | | | | $\beta_1$ | p>\|z\| | OR | $\beta_{1Std}$ | $\beta_2$ | p>\|z\| | OR | $\beta_{2Std}$ |
| 01-MAT | 1,581 | -1027.99 | 0.056 | 0.142*** | 0.005 | 1.153 | 0.238 | 1.105*** | 0 | 3.020 | 0.542 |
| 02-PHY | 2,921 | -1927.94 | 0.033 | 0.118*** | 0.001 | 1.125 | 0.230 | 0.833*** | 0 | 2.300 | 0.386 |
| 03-CHE | 1,649 | -911.78 | 0.192 | 1.768*** | 0 | 5.857 | 1.449 | 1.039*** | 0 | 2.828 | 0.517 |
| 04-EAR | 831 | -518.32 | 0.091 | 0.187*** | 0.002 | 1.206 | 0.380 | 1.366*** | 0 | 3.918 | 0.652 |
| 05-BIO | 4,554 | -2780.13 | 0.099 | 0.770*** | 0 | 2.160 | 0.845 | 0.722*** | 0 | 2.059 | 0.349 |
| 06-MED | 6,689 | -3921.52 | 0.128 | 0.730*** | 0 | 2.076 | 1.381 | 0.548*** | 0 | 1.729 | 0.271 |
| 07-AVS | 1,443 | -873.64 | 0.126 | 0.347*** | 0 | 1.415 | 0.806 | 1.435*** | 0 | 4.201 | 0.717 |
| 08-CE | 735 | -424.72 | 0.137 | 0.198** | 0.01 | 1.219 | 0.554 | 1.743*** | 0 | 5.714 | 0.867 |
| 09-IIE | 2,962 | -1728.99 | 0.144 | 0.149*** | 0.006 | 1.161 | 0.262 | 1.867*** | 0 | 6.471 | 0.934 |
| 11-PP | 1,064 | -653.01 | 0.056 | -0.009 | 0.473 | 0.991 | -0.043 | 1.143*** | 0 | 3.137 | 0.568 |

*Dependent variable: habilitation outcome; method of estimation: logistic regression QMLE; $\beta$ = raw coefficient; $\beta_{Std}$= X standardized coefficient. OR= Odds Ratio (exp $\beta$). z = z-score for test of $\beta$=0; p>|z| = p-value for z-test; Statistical significance: \* p-value <0.10, \*\* p-value <0.05, \*\*\* p-value <0.01.*

We note that in any case, the general statistics for the fitting parameters (columns 3 and 4 of Tables 6 and 7) indicate a scarce capacity of these two variables to alone explain the outcome of the evaluations conducted by the committees. For this, in the next section we consider three other potential explicative variables. To do this we must restrict the analyses to incumbents alone, for whom we have detailed information concerning the candidates' academic career and research production.

**4.2 Determinants of the habilitation outcome for incumbents: social proximity vs scientific merit**

The influence of social proximity between applicant and evaluator has already been examined by Abramo et al. (2014b), concerning its effects on efficiency of the selection for career advancement in Italian universities. Along the same lines, we now investigate the influence of social proximity on the habilitation outcome for incumbents, distinguishing two types of candidate-evaluator links:
- institutional link: the candidate is from the same university as the evaluators, at least for one year in the period 2003-2012;
- professional link: the candidate has co-authored publications with his/her evaluators, in the same period.

We thus examine the careers of both the applicants and evaluators to determine if the applicant spent at least one year in the same university as his or her evaluators, over the period 2003-2012[12].

Instances of shared professional work can be objectively measured by the proxy of publications in co-authorship. To analyze the influence of such candidate-evaluator research collaborations on habilitation outcomes we selected the 2003-2012

---

[12] The data on careers were extracted from cercauniversita.cineca.it/php5/docenti/cerca.php, last accessed 01/07/2015.

publications indexed in WoS[13].

Finally, considering the importance of belonging to a domestic academic community in a context (as in Italy's) where the scientific sector (the CS) is effectively the sphere of governance for both the individual and the collective's interests, we consider a further variable that discriminates between the incumbents on faculty in the same CS as the committee, and those on faculty in other CSs[14]. Finally, for the analysis of the habilitation to full professor, we discriminate the applicants already on staff as associate professors and as assistant professors ("ricercatore", in Italian), hypothesizing that beyond scientific merit, the committees could have also more or less formally applied hierarchical criteria. Formally, for the analysis of habilitations to full professor, the statistical model is described:

$$logit(p) = \beta_0 + \beta_1 X_1 + \beta_2 X_2 + \beta_3 X_3 + \beta_4 X_4 + \beta_5 X_5$$

[2]

Where:
$logit(p) = log \frac{p(E)}{1-p(E)}$
$E = 1$ if the applicant was habilitated; otherwise 0;
$p(E)$ = probability of event E;
$\beta$ = generic regression coefficient;
$X_1$ = applicant's bibliometric performance;
$X_2 = 1$ if the applicant was affiliated with the same university as at least one member of the committee, at least for a year over the period 2003-2012; otherwise 0;
$X_3 = 1$ if the applicant co-authored at least one publication with at least one member of the committee over the period 2003-2012, otherwise 0;
$X_4 = 1$ if at 31/12/2012, the applicant officially belonged to the same CS as that for the committee; otherwise 0;
$X_5 = 1$ if the applicant is an associate professor at 31/12/2012; otherwise 0.

For the habilitations to associate professor, the model is identical except for the absence of variable $X_5$.

Table 7 presents the results from the regressions for the habilitations to full professor. The candidates' bibliometric performance has a greater impact than other regressors in five UDAs out of ten (Chemistry, Earth sciences, Biology, Medicine and Civil engineering) and is not significant in two (Mathematics and Agricultural and veterinary science). The candidate's belonging to the same CS as the committee had an impact that was significant in all the UDAs, and superior to the impact of all the other regressors in five UDAs (Mathematics, Physics, and Agricultural and veterinary science, Industrial and information engineering, Psychology and pedagogy).

The institutional proximity ($X_2$) had a significant impact in all the UDAs except Physics and Agricultural and veterinary science; professional proximity ($X_3$) does not show significant impact in Chemistry and Earth sciences.

---

[13] The bibliometric dataset used to identify co-authorships is extracted from the Italian Observatory of Public Research (ORP), a database developed and maintained by the authors and derived under license from the Thomson Reuters WoS. Beginning from the raw data of the WoS, and applying a complex algorithm for reconciliation of the author's affiliation and disambiguation of the true identity of the authors, each publication (article, article review and conference proceeding) is attributed to the university scientist or scientists that produced it (for details, see D'Angelo et al., 2011).
[14] We recall that a candidate can apply for habilitation in different CSs, other than the one where he or she is officially categorized by the MIUR.

The analysis for habilitation to associate professor (Table 8) reveals a situation that is clearly different than the preceding one, even though there are some elements of continuity. The impact of bibliometric performance is significant in all the UDAs except in Physics. The candidate's belonging to the same CS as the committee has an impact that is significant in all the UDAs and higher than that of all the other regressors in Physics, Earth sciences, Industrial and information engineering, and Psychology and pedagogy. In the remaining UDAs, it is bibliometric performance that has a greater impact. Institutional proximity has a significant impact in only three UDAs and professional proximity in all UDAs but two.

In extreme summary and even though there are evident differences between the UDAs, the determinants seem to have different weights according to whether the habilitation is to associate or to full professor: in the first case the bibliometric performance seems to have had a much greater weight than for habilitation to full professor. In general, belonging to a CS different from that of the committee seems to penalize the candidates, as does the academic rank of assistant professor in the habilitations to full professor.

To give a quantitative idea of the weight of the single variables on the probability of habilitation of a candidate, we give an example from UDA 9-Industrial and information engineering. We will compare two candidates with bibliometric scores ($X_1 = 3$) three times greater than the benchmark: the first officially belongs to the CS where he or she is applying, the second does not. We begin with the habilitation to associate professor. A simulation conducted on the logistic regression results indicates that the "non-member" of the CS has a 49% probability of being habilitated. The individual's chances rise to 52% if he or she shows a link of institutional proximity to at least one of the committee members; to 56% if the person presents a link of professional proximity, and to 59% if there are both links. The "CS member" instead has a 79% probability of being habilitated and the probability rises by two points for each proximity link with the committee. Now we consider the habilitation to full professor. Both the candidates in the example are associate professors with a performance three times superior to the benchmark: the first, on faculty in a CS different than the one where the individual is applying, has a 15% probability of being habilitated, rising to 19% in the presence of either social link, and to 22% in the presence of both. The second individual, belonging to the CS of application, instead has a 41% probability of success, rising to 46% in the presence of a link of institutional proximity with the committee; to 48% in the presence of a professional link and to 53% in the presence of both. Finally, we analyze the other possible situation: an associate-professor and an assistant-professor candidate, both belonging to the same CS as the committee, both with a performance five times greater than the benchmark. For the habilitation to full professor, the assistant professor has a 75% probability of success, while the associate professor has an 87% probability.

*Table 7: Logistic regression results predicting outcomes for academic incumbents applying to full professor habilitation*

| | | | | Constant | | $X_1$ | | $X_2$ | | $X_3$ | | $X_4$ | | $X_5$ | |
|---|---|---|---|---|---|---|---|---|---|---|---|---|---|---|---|
| UDA | Obs. | Log pseudo likelihood | PseudoR$^2$ | $\beta_0$ | P>z | $\beta_{1Std}$ | P>z | $\beta_{2Std}$ | P>z | $\beta_{3Std}$ | P>z | $\beta_{4Std}$ | P>z | $\beta_{5Std}$ | P>z |
| 01-MAT | 727 | -449,45 | 0,094 | -2.301*** | 0 | 0.183 | 0.446 | 0.159** | 0.045 | 0.181** | 0.033 | 0.604*** | 0 | 0.426*** | 0 |
| 02-PHY | 706 | -426,63 | 0,108 | -1.658*** | 0 | 0.294** | 0.038 | 0.150 | 0.118 | 0.167* | 0.094 | 0.628*** | 0 | 0.355*** | 0 |
| 03-CHE | 537 | -308,67 | 0,152 | -3.179*** | 0 | 1.430*** | 0 | 0.310*** | 0.005 | 0.078 | 0.474 | 0.455*** | 0 | 0.095 | 0.362 |
| 04-EAR | 215 | -122,22 | 0,178 | -4.391*** | 0 | 1.035*** | 0.001 | 0.249* | 0.083 | 0.063 | 0.707 | 0.794*** | 0.003 | 0.551*** | 0.001 |
| 05-BIO | 1035 | -643,42 | 0,103 | -2.431*** | 0 | 0.705*** | 0 | 0.233*** | 0.001 | 0.302*** | 0 | 0.329*** | 0 | 0.282*** | 0 |
| 06-MED | 2192 | -1285,16 | 0,151 | -3.169*** | 0 | 1.077*** | 0 | 0.166*** | 0.001 | 0.333*** | 0 | 0.528*** | 0 | 0.389*** | 0 |
| 07-AVS | 506 | -284,86 | 0,154 | -2.643*** | 0 | 0.524 | 0.546 | 0.139 | 0.204 | 0.248** | 0.029 | 0.960*** | 0 | 0.087 | 0.389 |
| 08-CE | 286 | -148,54 | 0,246 | -4.072*** | 0 | 1.518*** | 0 | 0.345** | 0.020 | 0.415* | 0.091 | - | - | 1.065*** | 0 |
| 09-IIE | 1197 | -558,23 | 0,326 | -6.081*** | 0 | 1.157*** | 0 | 0.292*** | 0 | 0.196** | 0.024 | 1.375*** | 0 | 0.865*** | 0 |
| 11-PP | 263 | -119,95 | 0,307 | -6.072*** | 0 | 0.708* | 0.091 | 0.279* | 0.079 | 0.367** | 0.022 | 1.436*** | 0 | 1.100*** | 0.001 |

$X_1$= bibliometric performance; $X_2$=applic/evaluat. institutional proximity; $X_3$= applic/evaluat. coauthorships; $X_4$=application in the same CS of the applicant; $X_5$=applicant associate professor. Dependent variable: habilitation outcome; method of estimation: logistic regression QMLE; $\beta$ = raw coefficient; $\beta_{Std}$= X standardized coefficient. z = z-score for test of $\beta$=0; p>|z| = p-value for z-test; Statistical significance: * p-value <0.10, ** p-value <0.05, *** p-value <0.01

*Table 8: Logistic regression results predicting outcomes for academic incumbents applying to associate professor habilitation*

| | | | | Constant | | $X_1$ | | $X_2$ | | $X_3$ | | $X_4$ | |
|---|---|---|---|---|---|---|---|---|---|---|---|---|---|
| UDA | Obs. | Log pseudo likelihood | PseudoR$^2$ | $\beta_0$ | P>z | $\beta_{1Std}$ | P>z | $\beta_{2Std}$ | P>z | $\beta_{3Std}$ | P>z | $\beta_{4Std}$ | P>z |
| 01-MAT | 948 | -600,88 | 0,078 | -1.427*** | 0 | 0.682*** | 0 | 0.004 | 0.959 | 0.425*** | 0 | 0.376*** | 0 |
| 02-PHY | 913 | -492,74 | 0,107 | -0.404** | 0.045 | 0.297 | 0.14 | 0.071 | 0.403 | 0.133 | 0.166 | 0.721*** | 0 |
| 03-CHE | 912 | -437,97 | 0,220 | -3.637*** | 0 | 2.138*** | 0 | 0.281*** | 0.004 | -0.047 | 0.602 | 0.633*** | 0 |
| 04-EAR | 292 | -162,06 | 0,139 | -1.796*** | 0.001 | 0.586 | 0.108 | 0.226 | 0.129 | 0.332* | 0.094 | 0.792*** | 0 |
| 05-BIO | 1674 | -1052,68 | 0,090 | -1.672*** | 0 | 0.907*** | 0 | -0.005 | 0.927 | 0.124**ì | 0.02 | 0.304*** | 0 |
| 06-MED | 2866 | -1566,03 | 0,209 | -2.973*** | 0 | 1.809*** | 0 | 0.006 | 0.899 | 0.393*** | 0 | 0.612*** | 0 |
| 07-AVS | 749 | -358,97 | 0,242 | -3.662*** | 0 | 2.405*** | 0 | 0.244** | 0.019 | 0.326*** | 0.004 | 0.842*** | 0 |
| 08-CE | 409 | -227,58 | 0,191 | -3.572*** | 0 | 1.096** | 0.022 | 0.047 | 0.703 | 0.527*** | 0.003 | 1.048*** | 0 |
| 09-IIE | 1490 | -706,36 | 0,280 | -3.372*** | 0 | 1.108*** | 0 | 0.140* | 0.056 | 0.271*** | 0.001 | 1.365*** | 0 |
| 11-PP | 469 | -253,93 | 0,219 | -2.809*** | 0 | 0.258* | 0.081 | 0.334*** | 0.002 | 0.233** | 0.015 | 1.338*** | 0 |

$X_1$= bibliometric performance; $X_2$=applic/evaluat. institutional proximity; $X_3$= applic/evaluat. coauthorships; $X_4$=application in the same CS of the applicant. Dependent variable: habilitation outcome; method of estimation: logistic regression QMLE; $\beta$ = raw coefficient; $\beta_{Std}$= X standardized coefficient. z = z-score for test of $\beta$=0; p>|z| = p-value for z-test; Statistical significance: * p-value <0.10, ** p-value <0.05, *** p-value <0.01.

## 5. Discrimination and favoritism

The nationally governed competitions for faculty positions have come under frequent fire, and the Italian word "concorso" has gained international note for its implications of rigged competition, favoritism, nepotism and other unfair selection practices (Gerosa, 2001). Letters in prestigious journals such as The Lancet, Science and Nature (Garattini, 2001; Aiuti et al., 1994; Biggin, 1994; Amadori et al., 1992; Gaetani and Ferraris 1991; Fabbri, 1987), as well as entire monographs, (Perotti, 2008; Zagaria, 2007) continue to report on injustice in recruitment, including many cases that arrive in judicial proceedings. The habilitation competitions were conceived to contrast unfair selection practices, through the introduction of bibliometric indicators in the evaluation process.

In this section we attempt to quantify the numerosity of cases of probable discrimination and favoritism that may have occurred in the evaluation of candidates to associate and full professor habilitation. The operative definitions of discrimination and favoritism are based on the observation of bibliometric performance alone and vary according to three distinct scenarios:

Scenario 1 (restrictive)
- Favored candidates are those habilitated in a CS that: i) do not exceed any median; ii) are in the bottom 10% of the distribution of bibliometric performance for the candidates; and in any case iii) show performance that is less than that of at least one rejected candidate.
- Candidates subject to discrimination are those that were not habilitated that: i) exceeded all three medians; and ii) are in the top 10% of the bibliometric performance distribution for the candidates.

Scenario 2 (broad)
- Favored candidates are those habilitated that did not exceed all three medians.
- "Discriminated" candidates are those not habilitated that exceeded all three medians.

Scenario 3 (intermediate)
- Favored candidates are those habilitated that: i) did not exceed all three medians; and ii) had a bibliometric performance at least 20 percentile points less than that of a candidate that was rejected for habilitation in the CS.
- Candidates subject to discrimination are those that were not habilitated that: i) exceeded all three medians; and ii) had a bibliometric performance at least 20 percentile points greater than a candidate that was habilitated in the CS.

For the "intermediate" scenario, the threshold of 20 points of difference responds to a logic of compensation for potential qualitative differences in the profiles of the candidates, in other dimensions than those monitored by the three bibliometric indicators.

For ease of communication, in the following subsections we omit the adjective "probable" in discussing the discrimination and favoritism that are the subject of investigation. The objectives of the analysis is to identify those CSs where the committee's work would deserve further investigation, for purposes of remedying any inefficiencies in selection.

### 5.1 Analysis of cases of discrimination

Table 9 summarizes the analysis concerning cases of discrimination. Adopting the definition of first scenario, the candidates subject to discrimination are 6.8% of the non-habilitated to full professor and 5.9% of the non-habilitated to associate. The percentages rise to 30% and beyond under scenarios 2 and 3, with a peak of 38.6% in scenario 2 for the lower rank. Scenario 2 presents 71 CSs (almost 2/3 of the total 109) where the percentage of "discriminated" among non-habilitated to full professor is observed at over 30%; in the case of associate professors under scenario 2, there are 49 such CSs. For scenario 3 (intermediate), this same percentage of discriminated (30% among non-habilitated) affects 61 CSs for the habilitation to full professor and 45 for that to associate. Thus in general, the frequency of cases of discrimination appears highly relevant, and in every case is greater for the habilitation to full professor than for that to associate. This last observation is in line with what emerged in the regression analysis, where between the two ranks of habilitation there was a clearly different weight of the bibliometric performance on evaluation outcomes.

*Table 9: Cases of probable discrimination in the three scenarios under analysis*

| Habilitation to rank of | Full professor | Associate |
|---|---|---|
| Total candidates | 11,427 | 24,429 |
| Accredited | 5,099 | 10,901 |
| Discriminated (% of total non-accredited): Scenario 1 | 6.8% | 5.9% |
| Discriminated (% of total non-accredited): Scenario 2 | 38.6% | 30.5% |
| Discriminated (% of total non-accredited): Scenario 3 | 35.1% | 29.0% |
| No. sectors with % discriminated > 10%: Scenario 1 | 24 (of 109) | 14 (of 109) |
| No. sectors with % discriminated > 30%: Scenario 2 | 71 (of 109) | 49 (of 109) |
| No. sectors with % discriminated > 30%: Scenario 3 | 61 (of 109) | 45 (of 109) |

Because candidates were allowed to seek habilitation in more than one CSs, and a large number of them did, it may be the case that failures, notwithstanding high bibliometric performances, cannot be regarded as discrimination, rather as inadequate fit of the candidate with the competition sector. We therefore repeated the above analysis to candidates seeking habilitation in one CS only. Results are reported in Table 10.

*Table 10: Cases of probable discrimination in the three scenarios under analysis (candidates seeking habilitation in one CS only)*

| Habilitation to rank of | Full professor | Associate |
|---|---|---|
| Total candidates | 4,717 | 11,134 |
| Habilitated | 1,588 | 4,061 |
| Discriminated (% of total non-accredited): Scenario 1 | 10.2% | 9.4% |
| Discriminated (% of total non-accredited): Scenario 2 | 49.5% | 39.3% |
| Discriminated (% of total non-accredited): Scenario 3 | 46.0% | 38.0% |
| No. sectors with % discriminated > 10%: Scenario 1 | 47 (of 109) | 49 (of 109) |
| No. sectors with % discriminated > 30%: Scenario 2 | 82 (of 109) | 70 (of 109) |
| No. sectors with % discriminated > 30%: Scenario 3 | 78 (of 109) | 68 (of 109) |

The habilitation rate is lower for the above candidates: 36.5% as compared to 44.6% for candidates to associate professors; and 33.7% as compared to 44.6% for candidates to full professors. Discrimination rates, consistently, increase for both academic ranks.

We now explore the discrimination rates for incumbents and outsiders for all candidate (Table 11) and for those applying to a CS only (Table 12). Results show higher discrimination rates for candidates from outside the academia when considering all candidates, but no differences when considering candidates to one CS only.

*Table 11: Cases of probable discrimination for incumbents and outsiders in the three scenarios under analysis*

|            | Full professor |          | Associate professor |          |
|------------|----------------|----------|---------------------|----------|
|            | Incumbents     | Outsiders| Incumbents          | Outsiders|
| Scenario 1 | 5.5%           | 8.7%     | 5.1%                | 6.3%     |
| Scenario 2 | 35.7%          | 43.2%    | 27.7%               | 32.0%    |
| Scenario 3 | 31.9%          | 40.1%    | 26.3%               | 30.4%    |

*Table 12: Cases of probable discrimination for incumbents and outsiders (candidates seeking habilitation in one CS only)*

|            | Full professor |          | Associate professor |          |
|------------|----------------|----------|---------------------|----------|
|            | Incumbents     | Outsiders| Incumbents          | Outsiders|
| Scenario 1 | 9.4%           | 11.1%    | 10.3%               | 9.1%     |
| Scenario 2 | 50.0%          | 49.0%    | 42.7%               | 38.0%    |
| Scenario 3 | 45.9%          | 46.1%    | 41.1%               | 36.8%    |

With reference to all candidates, the data concerning the discrimination rates, when disaggregated by competition sector, permit identification of the CSs with the greatest concentration of critical cases. In this regard, Table 13 lists the first 10 sectors by concentration of discriminated candidates among the total of non-habilitated, for each of the three scenarios. We observe a massive presence of sectors under UDA 9 (Industrial and information engineering): from three to five sectors for the higher-rank habilitation, depending on the scenario, and from three to four sectors for the lower-rank habilitation. Under the most restrictive scenario the minimal incidence of discriminated candidates in these 10 CSs is 13.5% (09/E4-Electric and electronic measurement systems); the maximum is 33.3% (09/B2-Industrial and mechanical plant), for the higher-rank habilitations; for the lower-rank habilitations the interval is narrower, from a minimum of 10.9% in 01/A1 (Mathematical logic and complementary mathematics) to a maximum of 18.8% in 07/H4 (Clinical veterinary and veterinary pharmacology). In the other scenarios the incidence of discriminated candidates in the total of non-habilitated rises notably: for the higher-rank habilitation, the incidence is never less than 50%, with peaks of 80% in 08/A1 (Hydraulics, maritime construction and hydrology) and 09/C1 (Fluid machines, energy and environmental systems). For the lower-rank habilitation the incidence of discriminated candidates under scenarios 2 and 3 is never less than 45%, with peaks of 60% (under scenario 2, for 04/A3-Applied geology and once again for 08/A1).

The analyses of the individual scenarios indicate that the incidence of cases of discrimination is still greater in the habilitation to full professor but similar between the two ranks in terms of the CSs involved. The Spearman coefficient of correlation between the associate and full professor ranks for the lists of CSs is: 0.630 for scenario 1; 0.683 for scenario 2; and 0.671 for scenario 3. Observing Table 13, concerning scenario 1 we can see that five CSs result as present in both the lists for the associate and the full professor ranks. The comparison between the lists relative to the three scenarios for the 109 sectors analyzed indicates an almost perfect correlation between the second and third scenarios (Spearman $\rho = 0.973$ for the habilitation to full professor and 0.929 for the associates).The correlation between the lists concerning scenarios 1 and 2 is lower but equally significant (Spearman $\rho$ 0.591 for full professor habilitation and 0.679 for associate), as also holds for scenarios 1 and 3 ($\rho = 0.662$ for full professor and ρ = 0.713 for associate).

*Table 13: List of the first 10 CSs for incidence of cases of probable discrimination on the total of non-habilitated (%), in the three scenarios analyzed*

| Habilitation to full professor | | | | | | Habilitation to associate professor | | | | | |
|---|---|---|---|---|---|---|---|---|---|---|---|
| Scenario 1 | | Scenario 2 | | Scenario 3 | | Scenario 1 | | Scenario 2 | | Scenario 3 | |
| 09/B2 | 33,3 | 08/A1 | 80.5 | 09/C1 | 80,0 | 07/H4 | 18,8 | 04/A3 | 60.6 | 01/A4 | 59,3 |
| 07/H2 | 18,2 | 09/C1 | 80.0 | 01/A4 | 61,0 | 09/C1 | 13,6 | 08/A1 | 60.0 | 11/E4 | 55,6 |
| 01/A1 | 17,4 | 09/A3 | 68.6 | 09/B1 | 60,0 | 01/A4 | 13,6 | 09/D1 | 59.8 | 04/A3 | 55,3 |
| 08/A4 | 16,0 | 09/D1 | 66.9 | 09/D1 | 58,8 | 08/A4 | 13,0 | 01/A4 | 59.3 | 09/D1 | 54,6 |
| 01/A6 | 15,8 | 08/B2 | 62.5 | 09/A3 | 57,1 | 09/E4 | 12,5 | 11/E2 | 56.9 | 09/B1 | 53,3 |
| 09/B3 | 15,8 | 04/A2 | 62.3 | 02/B3 | 57,1 | 09/A2 | 11,9 | 11/E4 | 55.6 | 11/E2 | 50,7 |
| 07/F1 | 14,8 | 09/E4 | 62.2 | 06/F4 | 54,6 | 02/B3 | 11,2 | 09/B1 | 53.3 | 06/F4 | 50,6 |
| 01/A4 | 14,6 | 04/A3 | 61.4 | 02/A1 | 54,2 | 07/F1 | 11,1 | 06/F4 | 52.6 | 02/B3 | 45,8 |
| 06/E3 | 13,9 | 05/A2 | 61.1 | 09/E4 | 54,1 | 01/A3 | 10,9 | 09/E4 | 46.9 | 09/C1 | 45,5 |
| 09/E4 | 13,5 | 01/A4 | 61.0 | 08/A1 | 53,7 | 01/A1 | 10,9 | 07/B2 | 46.3 | 09/E4 | 45,3 |

**5.2 Analysis of cases of favoritism**

Table 14 summarizes the analysis concerning cases of favoritism. Under the definition of the most restrictive scenario (scenario 1), the favored candidates are slightly more than 1% of the total habilitated to full professor, and 0.6% of those habilitated to associate. As a matter of fact the scenario seems extremely conservative: the percentage of those favored exceeds 10% of the habilitated in only two of the CSs for full professor competitions, and in no CSs for the associate habilitation.

The incidence of those favored rises notably in the other scenarios. In scenario 2 the percentage of favored candidates is 30.7% for the competitions to higher rank and 32.3% for the lower. The third scenario again shows an incidence of favored candidates that is greater for the associate rank (29.4%) than for full professor (27.8%). The number of CSs with a percentage of favored candidates greater than 30% of the habilitated ranges from a minimum of 38 (of 109 total), for the higher rank under scenario 3, to a maximum of 59 for the lower rank under scenario 2.

*Table 14: Probable cases of favored candidates under the three scenarios analyzed*

| Habilitation to | Full professor | Associate professor |
|---|---|---|
| Total candidates | 11,427 | 24,429 |
| Habilitated | 5,099 | 10,901 |
| Favored (% of total habilitated): scenario 1 | 1.1% | 0,6% |
| Favored (% of total habilitated): scenario 2 | 30,7% | 32,3% |
| Favored (% of total habilitated): scenario 3 | 27,8% | 29,4% |
| No. sectors with % favored > 10%: scenario 1 | 2 (of 109) | 0 (of 109) |
| No. sectors with % favored > 30%: scenario 2 | 50 (of 109) | 59 (of 109) |
| No. sectors with % favored > 30%: scenario 3 | 38 (of 109) | 51 (of 109) |

This aggregate analysis thus reveals a rate of favoritism that is generally lower than the rate of discrimination, and unlike the preceding analysis, greater for the competition to associate than to full professor rank.

The sectors with a higher rate of favoritism are listed in Table 15. Under scenario 1, concerning the habilitation to full professor, the maximum incidence of favored candidates is traced to 09/B2-Industrial and mechanical plant (14.3%), followed by 09/C1-Fluid machines, energy and environmental systems (11.4%). In the list of the first 10 we find two more CSs (09/G1-Automatics and 09/A3-Design and methods for industrial engineering, metallurgy) of UDA 09 (Industrial and information engineering). Under the other scenarios, still concerning habilitation to full professor, the CSs that appear most frequently are those that belong to UDA 6 (Medicine), although it is a Biology CS that top the lists (05/E2-Molecular Biology), with an incidence of 60.3% for scenario 2 and 53.8% for scenario 3. For the habilitation to associate professor, the sectors of UDA 9 are again at the top: 09/B1 (Production technologies and systems), 09/B2 (Industrial and mechanical plant) and 09/C1 (Fluid machines, energy and environmental Systems) appear in the lists under all three scenarios, as does also 02/B2 (Didactics and history of physics). A further three sectors of Industrial and information engineering (09/E2-Electrical energy systems, electrical convertors, machines and switches, 09/E4-Electric and electronic measurement systems and 09/G1-Automatics), appear in the lists for scenarios 2 and 3. At the same time, still for the habilitation to associate professor, CSs 01/A1 (Mathematical logic and complementary mathematics), 01/A6 (Operations research) and 07/F1 (Food sciences) appear in two of the three lists for the scenarios. The analysis of correlation between the rankings under the different scenarios reveals a strong correlation between scenarios 2 and 3 (Spearman $\rho = 0.963$), and weak correlation between scenarios 1 and 2 (Spearman $\rho = 0.241$) and between 1 and 3 (Spearman $\rho = 0.269$). Correlating the rankings of the 109 sectors for the associate and full professor competitions, we see a Spearman $\rho$ equal to 0.302 for scenario 1 and slightly higher for the other two (0.344 for scenario 2 and 0.327 for scenario 3), which indicates that in this case there is non-homogeneity in the behavior on the part of the committees.

*Table 15: List of the first ten CSs for incidence of favored candidates on total habilitated (%), under the three scenarios analyzed*

| Habilitation to full professor | | | Habilitation to associate professor | | |
|---|---|---|---|---|---|
| Scenario 1 | Scenario 2 | Scenario 3 | Scenario 1 | Scenario 2 | Scenario 3 |
| 09/B2  14.3 | 05/E2  60.3 | 05/E2  53.8 | 09/B2  7.3 | 09/G1  68.4 | 09/G1  64.9 |
| 09/C1  11.4 | 06/A1  58.6 | 06/A1  51.7 | 09/B1  5.3 | 09/E4  60.0 | 09/E4  56.7 |
| 06/F3  7.3 | 07/E1  52.8 | 06/D4  48.8 | 09/C1  5.1 | 09/B2  58.5 | 09/B2  53.7 |
| 09/G1  6.8 | 06/D2  50.0 | 06/E3  47.8 | 02/A2  3.6 | 02/B2  57.9 | 07/F1  53.6 |
| 01/A5  6.3 | 06/D4  49.6 | 11/E2  47.1 | 01/A1  3.5 | 07/F1  53.6 | 09/B1  52.6 |
| 04/A4  5.9 | 06/E3  47.8 | 09/B3  46.7 | 02/B2  3.3 | 09/B1  52.6 | 01/A6  52.2 |
| 06/A3  5.6 | 11/E2  47.1 | 06/D2  44.6 | 02/A1  3.2 | 09/C1  52.5 | 09/C1  50.8 |
| 01/A4  4.8 | 09/B3  46.7 | 07/E1  44.4 | 07/H1  3.2 | 01/A6  52.2 | 02/B2  50.7 |
| 02/A1  4.7 | 03/B2  46.4 | 11/E1  42.6 | 02/B3  3.1 | 09/E2  52.2 | 01/A1  49.1 |
| 09/A3  4.4 | 05/F1  46.2 | 03/D1  40.6 | 06/A2  2.9 | 09/A3  49.5 | 09/E2  47.8 |

## 6. Discussion and conclusions

The publication of the results from the first Italian call for scientific habilitation to associate and full professor provides the elements for a first evaluation of the effects of the most recent of a long series of reforms attempted in this country, in the hopes of contrasting favoritism and discrimination in academic recruitment.

The analyses conducted reveal very clearly that for candidates already on staff as assistant or associate professor, and aspiring to career progression, it is sufficient to have a lower scientific performance than that of the candidates external to the academia, in order to have the same probability of success. For the incumbents, the same can be said of the candidates classified in the sector (CS) of the competition, compared to those from other CSs; and again, in the habilitation to full professor, for the associate professors compared to the assistant professors. The deeper investigation of the incidence of possible cases of discrimination and favoritism, shows that the first appears more extensive than the second, and certainly so for the habilitation to full professor. The stratification of the data by UDA shows significant differences: in Mathematics, the bibliometric performance seems to have had relatively scarce weight on the results of the habilitation, certainly for the competitions to full professor. On the opposite front, in the Chemistry sectors the results seem to converge with those predictable on the basis of a purely bibliometric evaluation. The Industrial and information engineering UDA has the highest concentration of CSs with greater incidence of possible cases of discrimination and favoritism. The phenomenon of favoritism also seems to have affected various CSs of Medicine.

The results of the analyses could give credit to two interpretations: the first is that the model of analysis based on bibliometric indicators, being unable to embed all the dimensions for evaluation of scientific merit, would have a systematic bias of evaluation in favor of candidates external to the academia or the CS, or of lower academic rank. Such an interpretation would be in spite of good sense, which would lead one to hold that there is a strong correlation between research performance and other dimensions of scientific merit (capacity to attract funds, positions on editorial boards of international journals, management of R&D, activities in technology transfer, etc.).

The other possible interpretation is that the cases detected denote actual phenomena of discrimination and favoritism, as confirmed by the high numbers of legal claims accepted by the Administrative Courts, and in line with the findings from other studies on Italian university competitions (see Section 1). This second interpretation aligns with a diffuse vision that the Italian university system is a "closed shop", with barriers to entry from outside that become higher with increasing academic rank. Within the closed shop there would then be the many smaller closed shops of the CSs, which defend the tradition and purity of the field; but especially of the consolidated social networks, which present formidable obstacles to the entry of "external" candidates, even when these are themselves Italian professors. Finally, this would be a closed shop managed by means of profound, almost military hierarchies, where rank and seniority in rank carry significant weight in whatever decisions, and career progression is often merited because of respect for the social rules that the hierarchy imposes.

However, recognizing the limits of bibliometrics as instrument of research evaluation, and adopting the cautions that scientific argument imposes, we assume that the truth probably lies somewhere between the two interpretations. That is, we accept that in some cases the judgment of the committees served to correct erroneous results that would have otherwise arisen from purely bibliometric evaluation, while in others it served to favor one candidate or discriminate against others.

The analysis presented here also highlights that the habilitation procedure attracted a quantity of aspiring faculty on the same order of size as the current national university research staff. With the current restrictions on turnover, expected to last at least over the near future, the universities can only announce calls for a very low share of the posts that would be necessary to absorb all the habilitated candidates, and this will now certainly become a further source of significant tension surrounding the system, considering the impossibility of satisfying the career expectation of such a massive number of scientists. In addition there is the entropy imposed on the system by the enormous number of decisions pending while the administrative courts work through the claims from candidates who were refused habilitation, and who now appeal the work of the committees that judged them. After a second call for habilitations, in which the total applications were only a third of those for the first call, the government has suspended the entire procedure, while announcing the preparation of the latest round of reforms. The unofficial expectation is that certain elements of the current procedure will be changed, while the overall structure remains.

In light of the analyses conducted, we would suggest that in the upcoming competitions for habilitation, the requisites of merit for eligibility to membership in the committees be made more stringently dependent on performance, given that it is legitimate to hold that the inclination to discrimination and favoritism is less diffuse among top performers. Furthermore, we suggest that the composition of the committees be extended to scientists from outside the university sphere, at equal level to the full professors, specifically inviting the "research directors" of public research institutions. These would presumably have less vested interests in the university system and so be less inclined to support exchanges of favors. From the analysis of the minutes regarding candidates that were not habilitated in spite of a top-10% bibliometric score, the motivation by far most frequently cited for the refusal is the incomplete fit of the candidate with the competition sector. The increasing multi-disciplinary character of research, and the blurred boundaries of research fields contrast with the assumption of competition sectors as sealed chambers. This makes it insupportable that such a high number of top scientists would be judged "unsuitable" for such a motivation. For such cases of top-scientist refusals it should be made compulsory that the committees indicate the alternative sector for the candidate, which they observe as being better suited. In the case of habilitated candidates with scarce scientific performance, and of non-habilitated with high performance, the committees should also be requested to provide precise and public reasoning of their decisions. In fact for many of the committees, the minutes presented were reduced to the mere provision of the same text of reasoning for every candidate, with the substitution of several customizing adjectives in the decision. Finally, the autonomy awarded to the committees in defining the evaluation parameters and criteria should be flanked by a true assumption of responsibility (delegation and consequences), to disincentivize immoral comportments that cannot be tolerated in civil nations, let alone in public institutions devoted to higher education and the shaping of society itself.